# Insights into hydrogen-induced vacancy stability and creep in chemically complex alloys

Prashant Singh,[1,+,*] Yash Pachaury,[2,+] Aaron Anthony Kohnert,[2] Laurent Capolungo,[2] Duane D. Johnson[1,3]

[1]*Ames National Laboratory, US Department of Energy, Iowa State University, Ames, IA, USA, 50010*
[2] *Los Alamos National Laboratory, Los Alamos, NM, USA, 87545*
[3]*Department of Materials Science & Engineering, Iowa State University, Ames, IA, USA, 50011*

Hydrogen (H) content modifies the creep response of Fe-based alloys by altering thermodynamics of point-defects; here we identify the electronic-structure mechanism underlying this effect. Using spin-polarized first-principles calculations combined with a cluster dynamics formulation, we establish a general framework linking H-assisted vacancy stabilization to diffusion-mediated creep in BCC Fe, FCC Fe, and chemically complex FCC Fe-Cr-Ni alloys. H-vacancy binding analysis shows that H-stabilized vacancies form at low hydrogen content in BCC Fe but require much higher chemical potentials in FCC Fe and Fe-Cr-Ni alloys due to broader *d*-bands, electronic screening, and chemical disorder. Consequently, plastic deformation mediated by diffusive processes is expected to be far more strongly impacted in BCC Fe than in FCC alloys. These electronic-controlled trends determine steady-state vacancy populations and provide a symmetry-resolved microscopic basis for H-assisted creep in ferritic and austenitic steels.

***Keywords:*** *Electronic-structure, cluster-dynamics, point-defect, creep, structural alloys*

***** *Corresponding author(s): psingh84@ameslab.gov; ypachaur@lanl.gov*
**+** ***Equal contribution***

Vacancy-Hydrogen (V-H) complexes constitute a central atomistic motif by which H alters defect chemistry, transport, and mechanical response in metals [**1-6**]. This effect spans pure metals to complex engineering alloys and underpins H-assisted degradation phenomena like embrittlement, accelerated creep, void nucleation [**7,8**], and enhanced diffusion. Hydrogen stabilizes vacancies through the formation of complexes, increasing their equilibrium concentration, and modifies the effective vacancy mobility [**6,9**]. In diffusion- or climb-controlled creep, plastic flow is governed by point-defect thermodynamics and kinetics through the equilibrium vacancy concentration $c_v \propto \exp(-\beta E^{vac}_{form})$ and the vacancy-mediated diffusivity $D_v \propto \exp(-\beta E^{vac}_{mig})$, with $\beta = 1/k_B T$ ($k_B$ is Boltzmann constant, T is temperature in Kelvin) which together set the matter flux available for dislocation climb [**10**], while dislocation-core (Peierls) energetics dictate glide-climb competition [**11,12**].

Experiments consistently demonstrate that H degrades creep resistance of both ferritic/martensitic and austenitic steels [**13-16**], yet the governing mechanism remains unresolved despite decades of investigation. Conflicting interpretations persist, with Yokogawa *et al.* attributing creep degradation to decarburization effects [**13**], He *et al.* invoking H-modified carbide interfacial energetics and suppressed grain-boundary sliding [**14**], and Takazaki *et al.* identifying H-assisted vacancy production (HAVP) as the dominant mechanism [**15, 16**]. However, such interpretations need a unified description grounded in electronic-controlled defect energetics capable of addressing chemically complex alloys, heretofore entirely absent [**17-22**].

Here, we establish a general electronic-based mechanism linking HAVP to defect thermodynamics and creep in Fe-based structural systems. We examine the formation, stabilization, and equilibrium concentrations of V-H complexes in BCC Fe, FCC Fe, and chemically complex FCC Fe-Cr-Ni alloys as functions of hydrogen chemical potential and temperature. H-vacancy binding occurs at lower H content in BCC Fe, but only at much higher effective H-levels in FCC Fe and FeCrNi, indicating that diffusion-mediated plasticity is far more H-sensitive in BCC alloys. These trends provide a unified microscopic connection between H-modified defect energetics and creep susceptibility across different alloy chemistries and crystal structures.

To quantify these coupled electronic-defect-mechanical effects, we employed spin-polarized density-functional theory (DFT) calculations as implemented within Vienna ab-initio Simulation Package (VASP) [**23,24**] using preferred Perdew-Burke-Ernzerhof (PBE) functional [**25**] with a plane-wave energy cutoff of 400 eV. Structural and electronic relaxations (see **SI Fig. S1**) were done until energy and residual forces converged to $10^{-6}$ eV and $10^{-3}$ eV/Å, respectively. Brillouin-zone integrations were carried out using Monkhorst-Pack *k*-point meshes [**26**]. DFT-derived binding energy of vacancy complexes ($E_{bind}^{VH_n}$) were coupled to a cluster-dynamics (CD) framework describing the reaction-dominated evolution of H, V, and $VH_n$ complexes, with Arrhenius kinetics and H concentrations set by Sieverts' law using solubilities for BCC Fe and FCC FeCrNi [**27,28**]. See SI for granular method details [**29-35**].

The electronic origin of structural and thermodynamic change is illustrated in **Fig. 1a** for pure Fe FCC, placing magnetic response on an absolute temperature scale (0→1500 K) with key transitions from $\alpha \rightarrow \gamma$ and the Curie point ($T_C$) of BCC Fe [**36-39**]. With lattice expansion, Fe evolves from low-moment anti-ferromagnetic (AFM) or nonmagnetic (NM) states at small $a_{FCC}$, to a paramagnetic (PM) disordered-local-moment (DLM) regime, and to a stable ferromagnetic (FM) state with moments of 2-2.5 $\mu_B$. The volume dependence of the *d*-band in Fe places BCC Fe near a Stoner instability, where lattice expansion narrows the *d*-states and increases states at Fermi energy, $N(E_F)$, stabilizing FM. This magneto-volume sensitivity establishes the electronic background against which vacancies, hydrogen, and alloying elements modify local moments and defect energetics with expected intrinsic correlations with crystal structure. The proximity of FM, AFM, NM, and DLM solutions indicates competing magnetic states, with DLM serving as a proxy for the PM regime. While BCC Fe retains robust FM order up to $T_C$, the FCC phase requires sufficient lattice expansion to satisfy the Stoner criterion. These electronic descriptors therefore govern how magnetic states couple to defect energetics, which in turn influences vacancy-mediated diffusion processes.

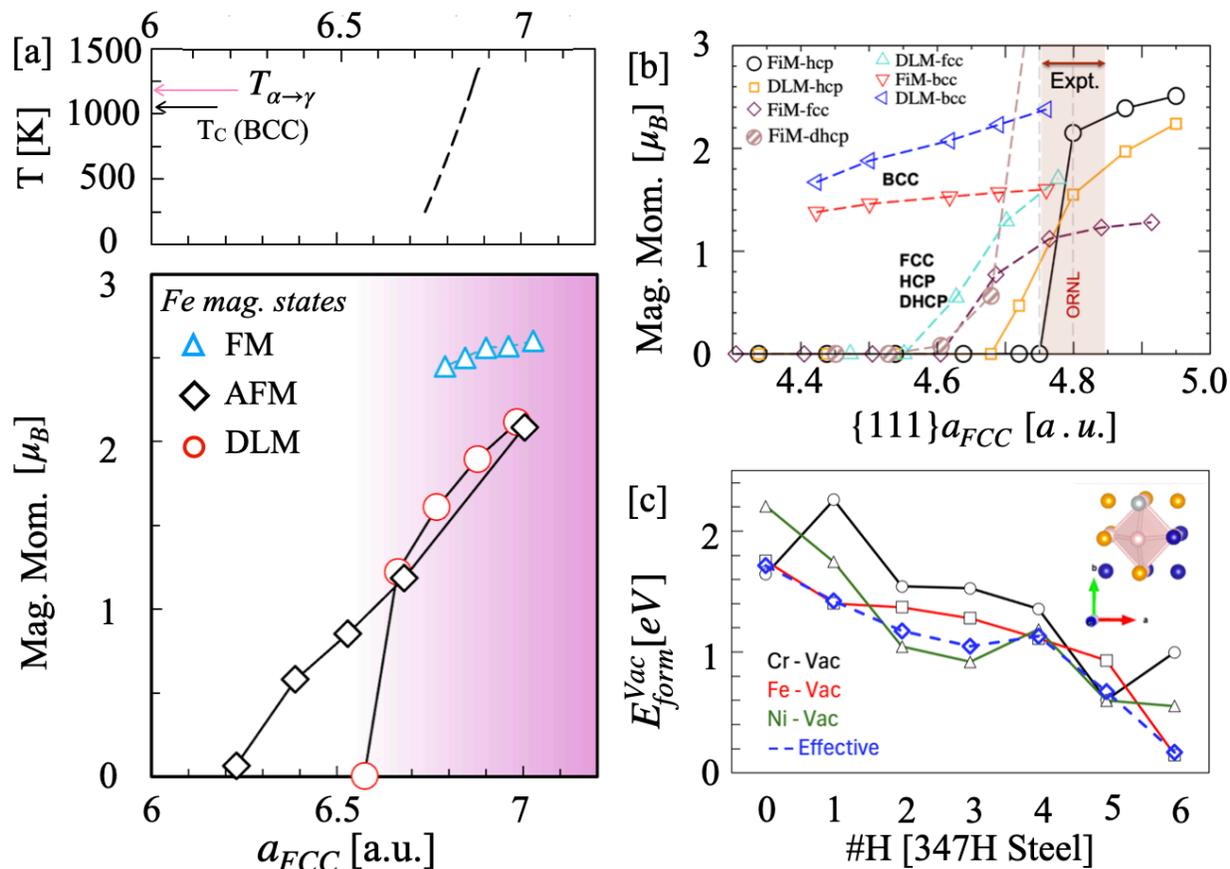

**Figure 1.** (a) Fe moment vs $a_{FCC}$ lattice constant for FM, AFM, and DLM states in pure Fe, showing a magneto-volume transition with moment quenching at small $a_{FCC}$ and stable FM moments (2-2.25 $\mu_B$) upon expansion. Top panel shows temperature scale marked by $T_C$ and $\alpha \rightarrow \gamma$ transition. For FCC FeCrNi, we show (b) moment vs $\{111\}$ $a_{FCC}$ for different crystal phases, highlighting earlier moment development in BCC than in FCC or HCP (shaded region is the $a_{FCC,Expt.}$), and (c) $E_{form}^{vac}$ vs trapped hydrogen showing strong H-induced stabilization. The shaded band (ORNL) indicates the measured lattice constant range, with possible error bars, for reference for actual Fe volumes.

Magnetization ($\mu_B$) vs. $\{111\}$ $a_{FCC}$ is plotted in **Fig. 1b** for multiple crystal structures and magnetic states, *i.e.,* ferrimagnetic (FiM) and DLM for FCC, BCC, hexagonal close-packed (HCP), and double-hexagonal close-packed (DHCP). The FCC FeCrNi for $a_{FCC,Expt}$ at 300 K is indicated by a shaded-vertical band [**40**]. For every phase, a critical lattice constant exists above which finite moments abruptly develop. The BCC branch is distinguished by the largest moment at a given lattice parameter and by sustaining magnetism down to smaller volumes, like BCC Fe. Notably, FCC or HCP require larger lattice expansions to satisfy the Stoner criterion. Observed $a_{FCC}$ fall near these thresholds for several phases, implying that modest volumetric or chemical perturbations can switch magnetic character. These abrupt onsets reflect a magneto-volume transition and explain why structural transformations ($\alpha \rightarrow \gamma$) in Fe strongly reshape magnetic stability, electronic structure, and thermodynamics.

To place vacancy formation and hydrogen stabilization on a thermodynamic footing, we use a grand-canonical defect formalism that treats H exchange with an external reservoir explicitly. For a VH$_n$ complex, the grand-canonical formation energy is defined as $\Delta G_f^{GC}(VH_n; T = 0K) = E_{form}^{vac} = E(VH_n) - E_{ref} + \mu_{Fe}$, where $E(VH_n)$ is the total energy of a supercell with one vacancy decorated by $n$ H-atoms, $E_{ref}$ is the energy of the pristine lattice with H atoms, and $\mu_{Fe}$ is the Fe chemical potential.

In **Fig. 1c**, the dependence is significant for $E_{form}^{vac}$, i.e., $\Delta G_f^{GC}(VH_n)$ at 0K, on local H content (H is at octahedral sites around the lowest-energy vacancy configuration). The formation energy in FeCrNi (dashed-blue line) is based on the partition of energy due to type of vacancy introduced as $E_f^{eff} = -(1/\beta).\ln[\sum_{i=1}^{3} g(E_f^i) \exp(-\beta E_f^i)]$, where probability density $g(E_f^i)$ is for a given vacancy microstate. Details on this equation are given in supplement (see SI Eq. 4 and **SI Fig. S5**. For all species, $E_{form}^{vac}$ decreases strongly with increasing H content, dropping from 1.5-3.0 eV at low occupancy to 0.5-1.0 eV by 4-6 H-atoms, demonstrating robust H-driven vacancy stabilization. Here, Cr-centered vacancies are most resistant at low H, while Fe and Ni exhibit larger early $E_{form}^{vac}$ reductions. At higher H-loadings the energies converge with minor plateaus reflecting site-specific relaxations and H-H interactions. Hydrogen stabilizes $VH_n$ complexes and lowers $E_{form}^{vac}$, which increases the probability of enhanced vacancy concentration outside these complexes. Under H-charging, this promotes vacancy nucleation, stabilization, and clustering, contributing to HAVP. The alloy-specific behavior emphasizes microchemical control where Cr-rich regions are expected to suppress vacancy formation owing to their high $E_{form}^{vac}$ except when surrounded by 5 H-atoms around Cr-site, where the formation energy is comparable to Fe and Ni vacancies. Moreover, Fe and Ni- centered vacancies dominate the bulk population owing to their low $E_{form}^{vac}$ and high Fe-content in 347H (17-20 at.% Cr, 9-13 at.% Ni, balance Fe) with the lower $E_{form}^{vac}$ of Ni-vacancies contributing more to the effective formation energies, evident in **Fig. 1c**. These results (with **Figs. 1a,b**) highlight the interplay between magnetic, structural, and defect energetics.

**Figure 2a** isolates crystal-symmetry effects by comparing $E_{form}^{vac}$ in BCC and FCC Fe. Both phases have similar baseline $E_{form}^{Vac}$ (1.5-2.2 eV) in the absence of H, but H produces a sharply divergent response in BCC, where $E_{form}^{Vac}$ collapses with the first few H and reaches near zero by n = 5 – 6. Whereas in FCC, the decrease is gradual and $E_{form}^{Vac}$ remains several tenths of an eV at n = 6. The origin is electronic-driven by BCC's low coordination, open tetrahedral channels and narrow, anisotropic *d*-band favor directional *d-s* rehybridization, giving large early V-H (VH, $VH_2$) binding with anisotropic relaxations amplified by magneto-volume coupling, collectively driving vacancy collapse and bond softening. FCC's higher coordination, broader *d*-bands and more isotropic screening weaken low-order H binding and delay stabilization, although fully-saturated $VH_6$ states can still bind strongly via delocalized Fe-H hybridization. In FCC 347H, Fe and Ni control bulk vacancy energetics, Cr locally elevates $E_{form}^{Vac}$, and vacancies become energetically stabilized only after multiple H atoms accumulate at high H concentrations. These results tie magnetic stability and electronic structure to alloy chemistry and explain why BCC microstructures are intrinsically prone to early vacancy nucleation and VH$_n$ clustering, while FCC austenitic alloys only delay that fate. Importantly, key contribution enters via the difference between systems with and without H. At zero H-content (n = 0), this reduces to conventional $E_{form}^{Vac}$, comparable in FCC and BCC Fe (1.5-2.2 eV). The stabilization by H arises from the cumulative V-H binding energies: as H binds with the vacancy, each added H lowers the total energy relative to the H-free reference, where incremental binding energy is $E_{bind}^{VH_n} = E_{form}^{VH_{n-1}} - E_{form}^{VH_n}$ (see **SI Eq. 14**). When the sum of binding energies exceeds the intrinsic $E_{form}^{Vac}$ (accounting for external $\mu_H$), the effective formation energy becomes zero or negative, indicating thermodynamic instability toward VH$_n$ formation. This decomposition makes explicit that H stabilizes vacancies through cumulative binding.

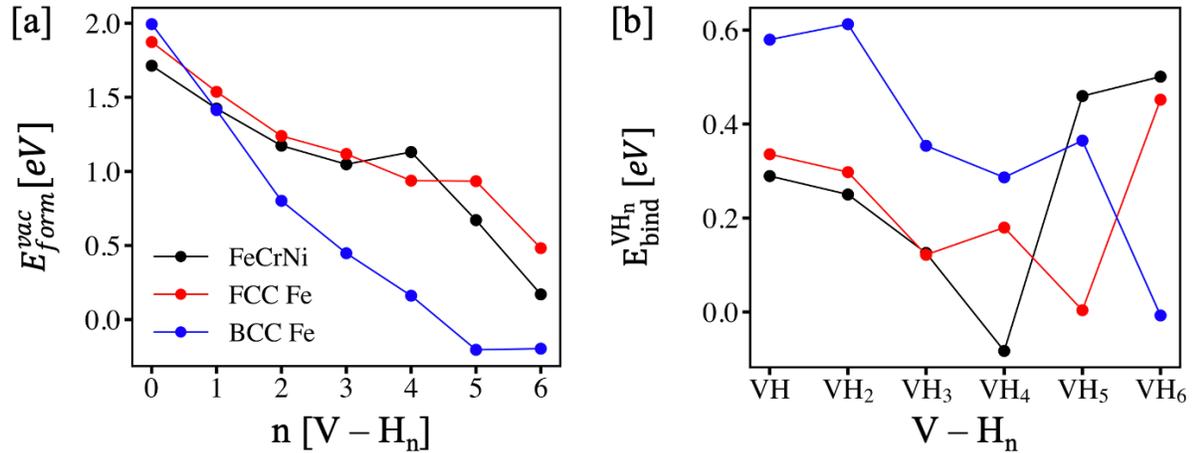

**Figure 2.** (a) $E_{form}^{vac}$, i.e., $\Delta G_f^{GC}(V - H_n)$ at 0K, vs trapped H number *n* in BCC Fe, FCC Fe, and FCC FeCrNi (347H). H rapidly collapses $E_{form}^{Vac}$ in BCC Fe, reaching near zero by *n*=5-6. In contrast, FCC Fe and FeCrNi show a more gradual reduction. (b) H binding energy ($E_{bind}^{VH_n}$) for $V-H_n$ (*n*=1-6) complexes, computed using **SI Eq. (12)**. Binding to low-order complexes ($VH, VH_2$) is strongest in BCC Fe, while FeCrNi shows enhanced stabilization of intermediate $VH_5 - VH_6$ complexes and FCC Fe exhibit the largest binding for the fully saturated $VH_6$ complex.

In BCC Fe, the first binding steps $E_b^{(1)}$ and $E_b^{(2)}$ in **Fig. 2b** shows they are large due to strong, anisotropic Fe-H bonding arising from narrow *d*-bands and low-coordination in a strongly magneto-volume-coupled Stoner FM, where local strain induced by hydrogen produces a substantial magneto-elastic energy gain. Consequently, $\sum_i E_i^{(b)}$ increases rapidly and satisfies $E_f(V) - \sum_{i=1}^n E_b^i(VH_i) \gtrsim E_f(VH_n)$ already for $n \approx$ 4-6, placing BCC Fe in a H-driven vacancy-formation regime with near-zero or negative $E_{bind}^{VH_n}$ and spontaneous VH$_n$ clustering. In FCC Fe, broader *d*-bands and more isotropic screening reduce early binding, so stabilization accumulates slowly, and vacancy collapse occurs only once the local hydrogen chemical potential and site occupancy are high enough to overcome the $E_{form}^{Vac}$ barrier. As discussed, Fe and Ni govern the average vacancy response whereas Cr locally increases the $E_{form}^{vac}$ in 347H. This combination suppresses early VH and VH$_2$ formation and does not support sustained cooperative stabilization at higher occupancies. Instead, once the vacancy-localized electronic states become saturated, the incremental $E_{bind}^{VH_n}$ turns negative (n ≥ 4), indicating destabilization of additional H due to vacancy-state saturation and the onset of H-H repulsion in the confined defect volume. This transition sharply increases stability of $VH_5$ and $VH_6$, consistent with enhanced vacancy retention at high H-contents. Overall, chemical disorder in the FeCrNi increases the formation energy of the thermally-activated vacancy (V, n = 0) and weakens the initial H-binding relative to BCC Fe. As a result, crossover from a thermally-activated vacancy regime (n = 0) to H-stabilized vacancy complexes (V-$H_n$, n ≥ 1) occurs only at substantially higher occupancies. Consequently, FeCrNi alloys are intrinsically less susceptible to H-assisted vacancy formation and clustering at low-to-moderate H activities.

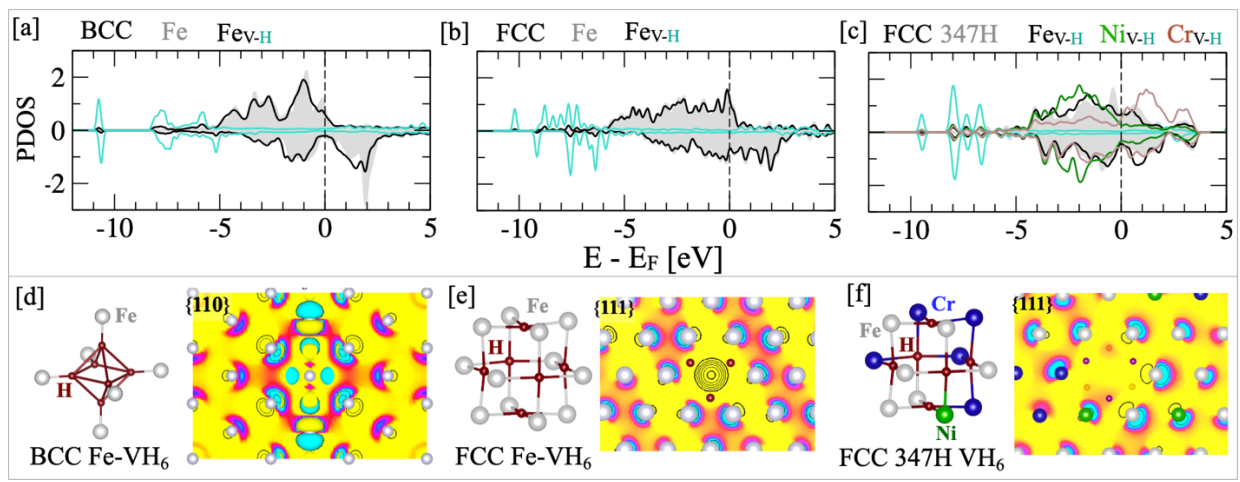

**Figure 3.** Partial density of states (PDOS) of (a) BCC Fe, (b) FCC Fe, and (c) FCC FeCrNi (347H) $VH_6$ complex (total DOS in grey is per Fe/atom including vacancy but without hydrogen), showing strong V-H induced *d*-band narrowing and electronic-state enhancement at the Fermi energy in BCC Fe, weaker redistribution in FCC Fe, and chemically inhomogeneous Fe/Ni/Cr-H hybridization in FeCrNi (more detail in **SI Fig. S2-S4**). 2D projected charge-density difference (Δ$\rho$) for $VH_6$ complex for (d) (110) BCC Fe, (e) (111) FCC Fe, and (f) (111) FCC FeCrNi, revealing directional rehybridization in BCC while isotropic but chemically heterogeneous screening in FCC systems. The charge density map was created with same parameter across system classes.

These thermodynamic trends are reflected in the electronic structure of BCC Fe, FCC Fe, and FCC 347H with and without V-H complexes (**Fig. 3**). In BCC Fe (**Fig. 3a**), vacancy formation reduces local coordination and redistributes Fe-*3d* spectral weight, producing sharper features near $E_{Fermi}$ indicative of enhanced electronic localization, rather than a simple increase in DOS at the Fermi energy. When H near vacancy, Fe-H hybridization modifies the occupied *d* states and drives charge redistribution toward the defect center (**Fig. 3d**). This charge redistribution is anisotropic, reflecting directional Fe-H bonding. The associated changes in local electronic structure alter magnetic moments in the vicinity of H, generating magnetic inhomogeneity and contributing to the defect energetics. The net effect is a strong, directionally localized electronic response that stabilizes V-H complexes relative to the bare vacancy. In contrast, FCC Fe (**Fig. 3b**) retains broader *d* bands characteristic of higher coordination and stronger metallic screening.

The PDOS changes upon V-H formation are more distributed and less localized, and the corresponding charge redistribution (**Fig. 3e**) is comparatively isotropic. Local Fe charges and magnetic moments vary within a narrow range, indicating that magnetic exchange remain coherent, and magnetically driven defect softening is limited. The chemical disorder further modulates this behavior in FCC 347H (**Fig. 3c**). With H added, element-centered environments exhibit distinct modifications. The spatially heterogeneous redistribution in **Fig. 3f** reflects the local chemical variability, where H stabilization becomes site dependent, producing locally preferred configurations rather than uniformly deep vacancy traps.

Charge-density difference maps (**Fig. 3d-f**) corroborate this electronic picture. In BCC Fe, H induces strongly anisotropic charge accumulation along Fe-H bonding directions, reflecting directional hybridization and a pronounced local electronic response to defect-induced strain. In FCC Fe, the charge redistribution remains largely isotropic and vacancy-centered, consistent with efficient metallic screening and the absence of strong directional localization. By contrast, FCC FeCrNi exhibits chemically modulated charge inhomogeneities, with H-induced redistribution strongly dependent on local Fe-Cr-Ni configuration. This electronic-driven view explains why BCC readily accumulate H at vacancies and extended defects, promoting the activation of diffusion-mediated plasticity, critical to creep deformation (apart from the dislocation creep regime), whereas FCC alloys mitigate directional trapping, producing weaker, spatially heterogeneous H stabilization.

These electronic trends directly manifest in the equilibrium populations of defects. The steady-state concentrations of H and $VH_n$ complexes in pure BCC and FCC Fe are shown in **Fig. 4**. These are computed via means of a CD model detailed in the SI. The equilibrium H concentrations in **Fig. 4a** indicate that FCC Fe can accommodate substantially higher H-content than BCC Fe, particularly at low temperatures. This is also evident from our DFT results that predict a higher solution enthalpy for BCC Fe (-0.058 eV/H) as compared to FCC Fe (-0.055 eV/H). The concentration of $VH_n$ (n=1-2) complexes is higher or nearly equal to that of vacancies at high $H_2$ gas pressure and at low-to-intermediate temperatures in BCC Fe, evident from **Fig. 4b**. In contrast, the concentration of the complexes in FCC Fe is significantly smaller than the concentration of the vacancies, seen in **Fig. 4c** at $p_{H2}$ of 1 bar. This is further evident from the apparent vacancy supersaturation defined as the relative increase in the net vacancy content in H-charged material with respect to the thermal-vacancy concentration. The apparent vacancy supersaturation is shown in **SI Fig. S6** while the role of $p_{H2}$ in modulating the concentration of vacancies in pure BCC and FCC systems is elucidated in **SI Fig. S7**.

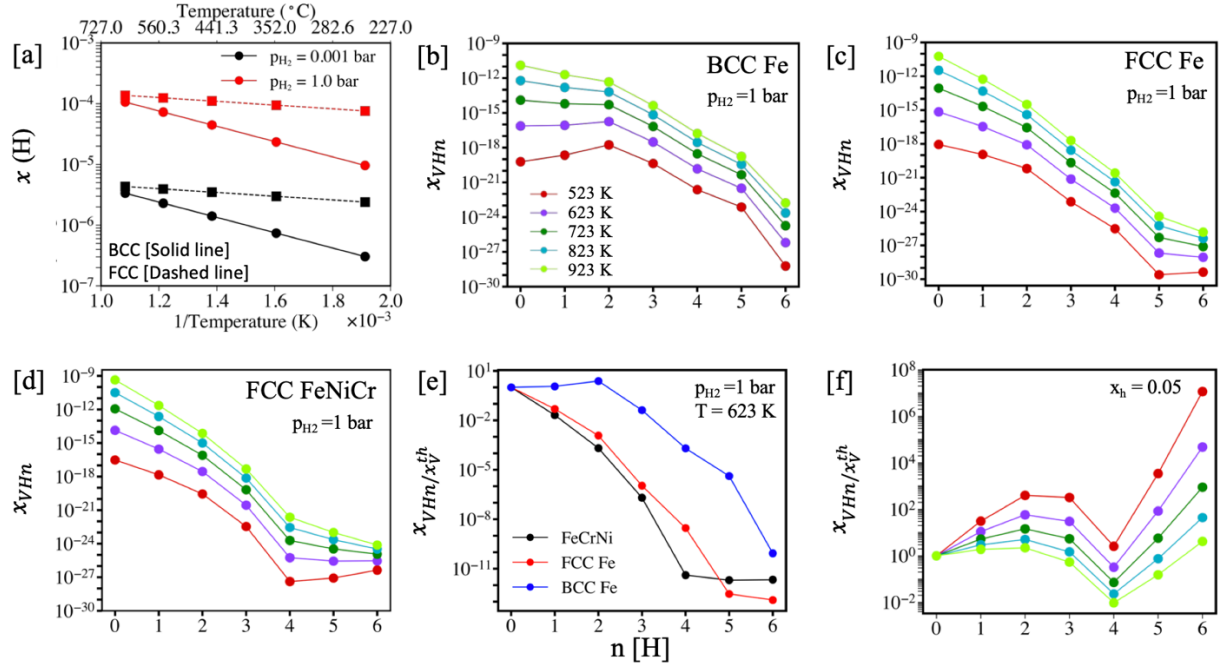

**Figure 4.** (a) Equilibrium H content in BCC and FCC lattices from Sieverts' law (**SI Eqs. 5-6**). (b-d) Steady-state concentrations of $VH_n$ (n=0-6) complexes in BCC Fe, FCC Fe, and FCC FeCrNi at $p_{H2}$ = 1 bar. In BCC Fe, VH and $VH_2$ compete with or exceed vacancies at low-intermediate temperature, whereas vacancies dominate in FCC Fe and FeCrNi. (e) Defect concentrations normalized to the thermal vacancy density at T = 623 K and $p_{H2}$ = 1 bar, highlighting the role of crystal structure and electronic screening. (f) FeCrNi at fixed H fraction (0.05), showing vacancy supersaturation stabilized as $VH_n$ complexes with $VH_6$ persisting at temperature below 823 K.

Like FCC Fe, vacancies are the dominant species among all $VH_n$ complexes for FCC FeCrNi alloys as seen in **Fig. 4d** across the spectrum of H-partial pressures ($p_{H2}$ = 0.001-1 bar) and temperatures (T=523-923 K) considered. Concentration of the complexes with respect to the thermal vacancy concentration reveals that it is smaller than the thermal vacancy concentration in FeCrNi as compared to FCC and BCC Fe (**Fig. 4e**). In other words, introducing H around a vacancy in FeCrNi is thermodynamically less favorable than pure FCC Fe case owing to the chemically complex environment. Nevertheless, the binding energy increases for $VH_5$ and $VH_6$ complexes which is manifested only at very high H concentrations in the lattice where H does lead to the excess vacancy concentration in the FeCrNi. **Figure 4f** shows one such case where the initial H concentration is taken as 0.05 ($n_H/H$). Such high concentrations occur under electrochemical charging or very high $p_{H2}$ (order of GPa). For such a case, we observe a significant increase in the concentration of the complexes as compared to vacancies especially at low-to-intermediate temperatures. Under these conditions, $VH_6$ complexes become dominant due to large binding energies of H in $VH_6$ (Fig. 2b). Moreover, as the temperature increases, the dissociation of the complexes also increases. Analytically, the concentration of any complex at steady state can be written in terms of the concentration of H and vacancies as $x^{ss}_{VH_n} = x^{th}_v (x^{ss}_H)^n \exp\left(\sum_{i=1}^{n} E^{VH_i}_{bind}/k_B T\right)$, which clearly demonstrates an inverse temperature dependence for the concentration of the complexes.

From a DFT perspective, the abrupt kink at $n = 4$ in **Fig. 4f** reflects a crossover in the dominant V-H complex. For $n \leq 3$, H occupies the most favorable sites around the vacancy, leading to smooth changes in calculated complex energies. Site-resolved energetics show that the four lowest-energy positions define a quasi-planar coordination shell. Occupation beyond this shell alters the local magnetic and elastic response of the vacancy, reducing the incremental stabilization at n = 4. Subsequent rearrangement partially restores binding, producing the observed non-monotonic trend. Notably, chemical disorder further amplifies this effect, as Fe-, Cr-, and Ni-centered vacancies respond differently at this occupancy, causing the Boltzmann-weighted average to become dominated by the most favorable local environments. This behavior reflects the combined effects of $E_{form}^{Vac}$ and elemental abundance. Cr-centered vacancies, despite occasionally strong H binding, contribute minimally due to their high $E_{form}^{Vac}$ under the presence of H atoms and low alloy concentration. On the other hand, Ni-centered vacancies possess the lowest $E_{form}^{Vac}$ for 2, 3, and 5 H atoms around their O-sites and therefore exhibit the strongest susceptibility to H-stabilization on a per-site basis. Lastly, Fe-centered vacancies dominate the VH, VH$_4$ and VH$_6$ population due to their much higher abundance along with significantly higher atomic fraction in 347H, thereby governing the averaged vacancy response observed in **Fig. 1c**. Consequently, the equilibrium VH$_n$ concentrations reflect a balance between H and V availability and H-binding strength. **Figures 2b & 4b-f** are therefore coupled via equilibrium thermodynamics, with the total VH$_n$ population given by the chemical environments incorporating $E_{form}^{Vac}$, H-binding energies, temperatures, H partial pressures and elemental fractions.

Finally, we assess the impact of HAVP on diffusion-dominated creep response of H-charged steels. Mukherjee-Bird-Dorn constitutive relationship has been widely utilized to correlate the steady state creep rates to material and microstructure parameters, such as diffusion coefficients, grain size, and elastic modulus [**41**]. Using this relationship, the steady-state creep rate in the diffusion dominated regime scales with the vacancy self-diffusivity, i.e. $\dot{\varepsilon}_{ss} \propto \exp[-\beta(E_{form}^{vac} + E_{mig}^{vac})]$, assuming other parameters, such as stress state, grain size, dislocation density, and structure factor to be fixed. **Fig. 5** shows the acceleration of steady-state creep rates in H-charged vs. H-free BCC Fe, FCC Fe, and FCC FeCrNi as a function of H partial pressures and temperature, using diffusion coefficients listed in **SI Table 1.** The creep acceleration is defined as the ratio of the steady-state creep rates in H-charged steel ($\dot{\varepsilon}_{ss}^{H}$) vs H-free steel ($\dot{\varepsilon}_{ss}^{noH}$), i.e., $\dot{\varepsilon}_{ss}^{H}/\dot{\varepsilon}_{ss}^{noH} = \sum_{n=0}^{6} D_{VH_n} c_{VH_n}^{ss} / D_V c_V^{th}$. High H partial pressures ($p_{H2}$) at low to intermediate temperatures increase the steady-state creep rate by up to 14% in BCC Fe (**Fig. 5a**), whereas it has a negligible effect on creep in FCC Fe (**Fig. 5b**) and FeCrNi (**Fig. 5c**). The net impact of vacancy supersaturation is limited by the slow migration kinetics of VH$_n$ complexes. Molecular dynamics studies indicate sluggish migration of VH$_1$ complex in BCC Fe [**42**], while migration rates for higher-order complexes are assumed here. Consequently, the trends in **Fig. 5** should be interpreted with caution, as H-lubrication effects like those reported by Du et al. [9] for Cu and Pd could enhance $VH_n$ mobility and further accelerate creep. For FCC Fe and FeCrNi, significantly higher H concentration in the lattice is expected to alter their creep response indicating austenitic steels are less susceptible to creep degradation due to HAVP in the diffusion dominated regime. Moreover, these results highlight the need for quantitative determination of $VH_n$ migration kinetics to fully elucidate H-assisted creep in Fe-based alloys.

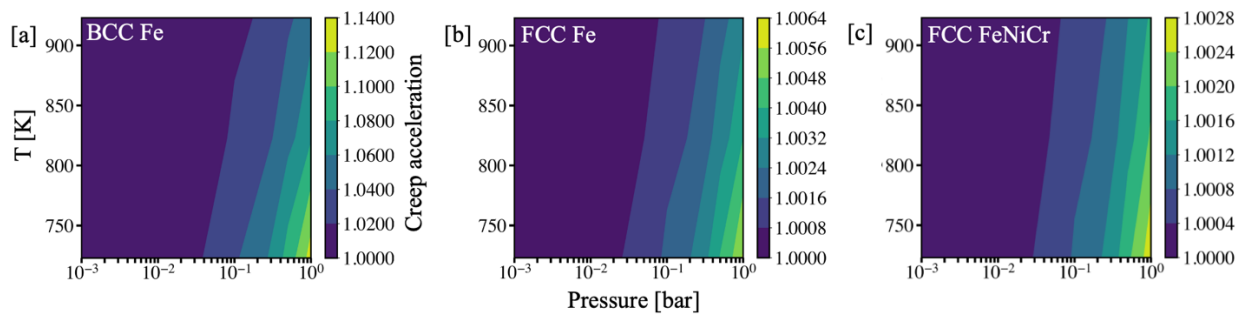

**Figure 5.** Creep acceleration ($\dot{\varepsilon}_{ss}^{H}/\dot{\varepsilon}_{ss}^{noH}$) for (a) BCC Fe, (b) FCC Fe, and (c) FCC FeCrNi follows the trend of steady-state $VH_n$ concentrations, but its effect is not 1:1 due to the slow kinetics of these complexes assumed in this work.

In summary, H-assisted vacancy formation in Fe-based materials emerges from the coupled effects of lattice symmetry, electronic-structure, magnetism, alloy chemistry, and defect kinetics. BCC Fe is intrinsically susceptible due to reduced coordination and lattice symmetry driven *d*-band narrowing, which enhances directional Fe-H hybridization that rapidly lowers vacancy formation energetics, and increases vacancy supersaturation relevant to diffusion-controlled creep. In contrast, FCC lattices suppress this instability through broader *d*-bands and efficient electronic screening, leading to weaker trapping and negligible H-induced creep acceleration. In FCC 347H, chemical disorder further modulates this behavior where Cr destabilizes low-occupancy V-H complexes, while Fe and Ni stabilize them only at higher chemical potentials. Using DFT and CD, this work establishes a unified electronic-based picture connecting H-modified vacancy thermodynamics to diffusion-mediated creep, explaining why austenitic steels mitigate but do not eliminate H-driven vacancy production and the potential role of slow $VH_n$ kinetics in delaying H-assisted degradation in structural alloys.

This work was supported by the U.S. DOE Office of Fossil Energy and Carbon Management through the eXtremeMAT (XMAT) program under Crosscutting Technology High-Performance Materials Research Program. YP, AAK, and LC also acknowledge support from Laboratory Directed Research and Development Program of Los Alamos National Laboratory under Project No. 20260042DR. The Ames National Laboratory is operated for the U.S. DOE by Iowa State University under contract DE-AC02-07CH11358. Los Alamos National Laboratory is operated by Triad National Security, LLC, for the National Nuclear Security Administration of the U.S. Department of Energy (Contract No. 89233218CNA000001).